# Parallel MRI at microtesla fields

Vadim S. Zotev*, Petr L. Volegov, Andrei N. Matlashov, Michelle A. Espy,
John C. Mosher, Robert H. Kraus, Jr.

*Los Alamos National Laboratory, Group of Applied Modern Physics, MS D454, Los Alamos, NM 87545, USA*

**Abstract**

Parallel imaging techniques have been widely used in high-field magnetic resonance imaging (MRI). Multiple receiver coils have been shown to improve image quality and allow accelerated image acquisition. Magnetic resonance imaging at ultra-low fields (ULF MRI) is a new imaging approach that uses SQUID (superconducting quantum interference device) sensors to measure the spatially encoded precession of pre-polarized nuclear spin populations at microtesla-range measurement fields. In this work, parallel imaging at microtesla fields is systematically studied for the first time. A seven-channel SQUID system, designed for both ULF MRI and magnetoencephalography (MEG), is used to acquire 3D images of a human hand, as well as 2D images of a large water phantom. The imaging is performed at 46 microtesla measurement field with pre-polarization at 40 mT. It is shown how the use of seven channels increases imaging field of view and improves signal-to-noise ratio for the hand images. A simple procedure for approximate correction of concomitant gradient artifacts is described. Noise propagation is analyzed experimentally, and the main source of correlated noise is identified. Accelerated imaging based on one-dimensional undersampling and 1D SENSE (sensitivity encoding) image reconstruction is studied in the case of the 2D phantom. Actual 3-fold imaging acceleration in comparison to single-average fully encoded Fourier imaging is demonstrated. These results show that parallel imaging methods are efficient in ULF MRI, and that imaging performance of SQUID-based instruments improves substantially as the number of channels is increased.

*Keywords:* Parallel MRI; ULF MRI; SENSE; MEG; SQUID

## 1. Introduction

Parallel magnetic resonance imaging (parallel MRI) is based on simultaneous acquisition of magnetic resonance signals with multiple receiver coils, characterized by distinct spatial sensitivities. Imaging with a coil array generally offers several advantages. First, additional coils enlarge imaging field of view (FOV) by providing sensitivity in those regions where sensitivity of a single coil is low. Second, multiple coils yield higher imaging signal-to-noise ratio (SNR), because MRI signal from each voxel is acquired simultaneously by several coils. The combination of images from the individual coils may improve image quality substantially, especially if it takes into account sensitivity and noise properties of the coils. Third, an array of coils allows accelerated imaging, because the spatial encoding effect of multiple coils is independent of the gradient encoding mechanism of conventional Fourier MRI. This effect can be used to perform some portion of encoding normally done with the gradients, and thus reduce imaging time.

Parallel MRI had been used primarily for FOV and SNR improvement (see, e.g., [1-4]) until about ten years ago, when the focus of parallel imaging studies began to shift towards imaging acceleration. This transition was precipitated by the progress in MRI technology and the need for faster medical imaging. Accelerated image acquisition in parallel MRI is achieved at the expense of reduction in imaging SNR. It is usually realized by undersampling along the phase encoding direction(s) that would lead to reduced FOV and aliasing in conventional Fourier imaging. Parallel MRI reconstruction methods

─────
* Corresponding author. Fax: +1 505 665 4507.
  E-mail address: vzotev@lanl.gov (V.S. Zotev).

incorporate coil spatial sensitivities to generate full-FOV images without aliasing from the undersampled data acquired with multiple channels. These methods can be divided into *k*-space (such as SMASH [5]), image-domain (SENSE [6]), and hybrid (such as non-Cartesian SENSE [7]) approaches. They can be formulated and compared within the same theoretical framework [8,9]. Parallel imaging methods are essential in MRI applications requiring high temporal resolution, such as cardiovascular MRI [10], abdominal MRI [11], and functional MRI of the human brain [12]. Acceleration factors as high as 16 have been reported in vivo [11], and as high as 64 with phantoms [13]. Yet higher temporal resolution can be achieved if parallel imaging is combined with magnetic source localization by a large sensor array [14] used in magnetoencephalography (MEG) [15]. In clinical practice, 2- to 4-fold accelerations are more typical.

Parallel imaging methods have until recently only been applied in conventional high-field MRI. Magnetic resonance imaging at ultra-low fields (ULF MRI) is a new imaging approach that uses measurement fields in the microtesla range [16-27]. Broadening of the NMR signal linewidth is determined by absolute inhomogeneity of the magnetic field. For a fixed relative inhomogeneity, the absolute field inhomogeneity scales linearly with the field strength [16]. Because of this, ultra-low fields of modest relative homogeneity are highly homogeneous on the absolute scale, and very narrow NMR lines with high SNR are achieved [16,20,26]. This fact greatly simplifies coil design and makes it possible to construct ULF MRI systems [18,24] that are simple, inexpensive, portable, and patient-friendly. Imaging at ULF offers additional benefits, including minimized susceptibility artifacts [16], enhanced $T_1$ contrast [28], and possibility of imaging in the presence of metal [20,29]. Moreover, ULF MRI is compatible with MEG as discussed below.

Implementation of MRI at ultra-low fields encounters two problems: insufficient magnetization of a sample by microtesla fields, and low efficiency of Faraday detection at kilohertz-range frequencies. The first difficulty is typically resolved by using the pre-polarization technique [30]. In this approach, the sample is pre-polarized by a relatively strong (up to 100 mT and higher) magnetic field prior to each imaging step. The second problem can be solved if highly sensitive SQUID (superconducting quantum interference device) sensors [31] are used to measure NMR signals [32-35]. SQUIDs with untuned input circuits are typically employed [16-27], because their response is independent of frequency. Low-field images acquired using tuned SQUID pre-amplifiers [33], as well as Faraday detector coils [36,37], have also been reported, though the method of [16-27] appears to be more efficient. Despite these improvements, insufficiently high SNR remains a major limitation in present-day ULF MRI. The SNR can be improved by stronger pre-polarization, which makes the situation similar to that in conventional MRI with its quest for higher magnetic fields. The pre-polarizing field in ULF MRI, however, does not need to be very uniform, because no spin precession is measured during the pre-polarization.

One property of ULF MRI makes it particularly attractive: it can be easily combined with SQUID-based techniques for biomagnetic measurements, such as MEG [15] and magnetocardiography (MCG) [38]. It has been demonstrated by our group that ULF NMR signals, generated inside a human body, can be measured simultaneously with MEG [21] or MCG [22] signals using the same SQUID sensor.

Recently, we used a seven-channel SQUID system [23,24], specially designed for MEG and ULF MRI, to acquire the first images of the human brain at microtesla fields [25]. We also recorded auditory MEG signals during the same imaging session [25]. This result demonstrated feasibility of human brain imaging by microtesla MRI and showed that multichannel SQUID systems for combined MEG and ULF MRI of the brain are practical. Such systems can directly provide anatomical ULF MRI maps for MEG-localized neural sources. They can also greatly facilitate integration of MEG with high-field MRI and other imaging modalities, because ULF images (and thus MEG data from the same system) can be precisely matched to structural images provided by other methods.

Because MEG instruments include large arrays (typically hundreds) of SQUID sensors, parallel imaging techniques should be very efficient in ULF MRI. Parallel imaging is easier to implement at ULF than in conventional high-field MRI for two reasons. First, because ULF MRI relies on untuned SQUID detection, there is no need to minimize inductive coupling between pick-up coils. This means that any SQUID array, suitable for ULF MRI, can be used for parallel imaging. Second, because electric currents, induced inside a sample, diminish with decreasing frequency, noise properties of a ULF MRI system are essentialy independent of the sample. Therefore, channel sensitivities can be accurately determined from phantom measurements. Similar to conventional high-field imaging, multiple sensors can be used at microtesla fields to improve imaging FOV, SNR, and speed. Image distortions due to concomitant gradients, however, are more pronounced at ULF, and need to be corrected as FOV increases [39-41].

In this work, the first systematic study of parallel MRI at microtesla fields is reported. The seven-channel SQUID system [23,24] is used to acquire 3D images of a human hand, as well as 2D images of a large phantom. Improvements in imaging FOV and SNR by the sensor array are illustrated for the hand images, with effects of concomitant gradients corrected and noise propagation taken into account. Actual 3-fold imaging acceleration is achieved in the phantom experiment using SENSE method [6].

## 2. Materials and methods

*2.1 Instrumentation*

All experimental results, reported in this paper, were obtained using the seven-channel SQUID system for 3D ULF MRI and MEG [23,24], depicted schematically in Fig. 1. The system includes seven second-order SQUID gradiometers with 37 mm diameter and 60 mm baseline, characterized by magnetic field resolutions of 1.2…2.8 fT/√Hz at 1 kHz [24]. The gradiometers are installed parallel to one another inside a flat-bottom liquid helium cryostat. Their pick-up coils form a symmetric pattern shown in Fig. 1A with 45 mm center-to-center spacing of the neighboring coils. Each SQUID sensor is equipped with a cryoswitch [24] that protects it from transients caused by switching fields and gradients in ULF MRI experiments. All experiments are performed inside a magnetically shielded room. It should be noted that the shielded room is needed for MEG measurements only, and can be replaced with a less expensive RF screen for ULF MRI at kilohertz-range frequencies.

As mentioned in the introduction, the measurement field and encoding gradients for ULF MRI can be generated by simple and inexpensive coil systems. Schematic of our coil system is shown in Fig. 1B. The ultra-low measurement field $B_m$ is created along the Z axis by a pair of round Helmholtz coils, 120 cm in diameter. The strength of the $B_m$ field is approximately 46 µT, which corresponds to the proton Larmor frequency of about 1940 Hz. Three sets of thin rectangular coils, symmetric with respect to the system center, generate three gradients for 3D Fourier imaging (Fig. 1B). The pre-polarizing field $B_p$ in the present system is three orders of magnitude stronger than the measurement field $B_m$. It is produced by a cylindrical coil positioned below the sample. The vertical component of the $B_p$ field varies between 40 and 50 mT across the sample space.

The 3D imaging procedure, used in our experiments, is shown in Fig. 1C. Each imaging step begins with pre-polarization of a sample by the field $B_p$ during time $t_p$. The pre-polarizing field is then turned off rapidly, and the measurement field $B_m$ is applied. In the present set-up, the $B_p$ field is ramped down linearly in 6 ms, and the $B_m$ field is applied 3-4 ms later, by which time all transients, induced in the system coils by the $B_p$ pulse, are dissipated. The application of the measurement field $B_m$ perpendicular to the original direction of $B_p$ induces spin precession [24]. Imaging is then performed according to the standard 3D Fourier imaging protocol with gradient echo. Spin precession is phase encoded by two gradients, $G_z$ and $G_y$, during time $t_g$, and gradient echo is created by reversal of the frequency encoding gradient $G_x$. The echo signal is measured during the acquisition time $t_a$ with discrete sampling at 16 kHz. The sequence in Fig. 1C is repeated for all combinations of the selected $G_z$ and $G_y$ values, and a 3D image is reconstructed from the acquired data by 3D fast Fourier transform.

*2.2 Imaging parameters*

The seven-channel system depicted in Fig. 1 was used to perform ULF MRI of objects shown in Fig. 2. They include a human hand and a water phantom. The phantom was constructed using a polyethylene disc with flat-bottom holes (19 mm in diameter and 19 mm deep) drilled in the pattern shown (Fig. 2) with 22 mm center-to-center spacing. The holes were filled with tap water with experimentally determined transverse relaxation time $T_2^* \approx 2.8$ s. The mean relaxation time $T_2^*$ for the human hand was measured to be ≈ 120 ms.

In the human hand experiment, the following imaging parameters were used (Fig. 1C). The hand was pre-polarized for $t_p$=0.5 s. The gradient encoding and signal acquisition times were $t_g$=42 ms and $t_a$=84 ms, respectively. The frequency encoding gradient $G_x$ changed between ±94 µT/m (±40 Hz/cm). The phase encoding gradient $G_z$ had $N_z$=55 different values, that were equally spaced and symmetric with respect to $G_z$=0, with the maximum value $G_{z,max}$ = -$G_{z,min}$= 94 µT/m (40 Hz/cm). Similarly, $N_y$=9 values were selected for the gradient $G_y$, with the maximum value $G_{y,max}$ = -$G_{y,min}$= 47 µT/m (20 Hz/cm). These imaging parameters provided 3 mm × 3 mm × 6 mm resolution, with the 6 mm pixel size corresponding to the vertical dimension. A single scan of $k$ space included 495 measurements and required 5 min. To improve image quality, results of 12 consecutive scans were averaged. The total imaging time was about one hour. It should be noted, however, that 80% of this time was taken up by pre-polarization.

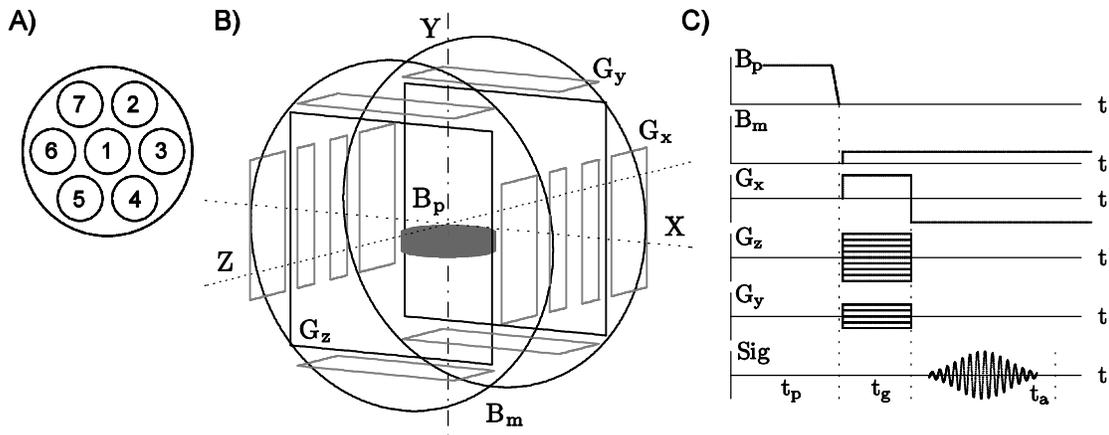

Fig. 1. The seven-channel system for 3D ULF MRI and MEG. (A) Positions of the SQUID channels inside the cryostat, (B) schematic of the coil system, (C) 3D Fourier imaging sequence with pre-polarization and gradient echo.

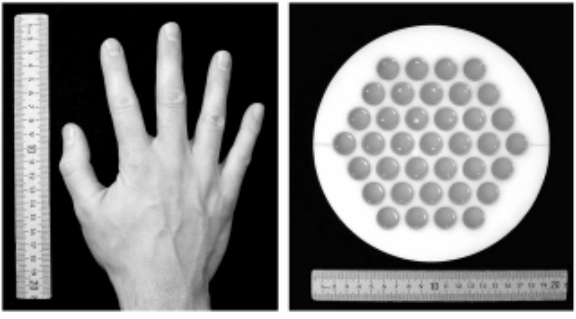

Fig. 2. Objects for imaging: a human hand and a water phantom.

The purpose of the 2D water phantom experiment was to demonstrate imaging acceleration by using parallel image reconstruction. Parameters of the imaging procedure, therefore, were selected to maximize SNR and allow image acquisition without signal averaging. They had the following values: $t_p$=4 s, $t_g$=250 ms, $t_a$=500 ms, $G_x$= ±23.5 µT/m (±10 Hz/cm), $N_z$=73, $G_{z,max}$ = -$G_{z,min}$= 23.5 µT/m (10 Hz/cm), $N_y$=1, $G_y$=0. This 2D imaging sequence provided 2 mm × 2 mm resolution in the XZ plane. Because of the long pre-polarization time required to polarize water, single-average acquisition of a full-FOV phantom image took about 6 min.

After each of the the two imaging experiments, sensitivity maps of the seven channels were acquired by imaging a large uniform water phantom placed under the cryostat instead of the studied object. The imaging resolution was the same as in the main experiment. The imaging parameters were also the same except for a smaller increment value and greater number of steps $N_z$ for the $G_z$ gradient to obtain a larger FOV in the Z direction. The pre-polarization time for the uniform water phantom was at least 2 s.

*2.3 Concomitant gradients*

The system of seven pick-up coils, shown in Fig. 1A, provides the imaging FOV in the horizontal XZ plane that is approximately 3 times as wide as the FOV of a single coil. Because image distortions caused by concomitant gradients grow quadratically with the distance from the imaging center, they become more pronounced as FOV increases. While correction of concomitant gradient artifacts is rather complicated in the general case [39,40], it can be simplified for the imaging parameters used in our experiments.

It has been shown [41] that image distortions due to concomitant gradients can be characterized by a dimensionless parameter $\varepsilon = L(G/B_m)$, where $L$ is the object size, $G$ is the imaging gradient strength, and $B_m$ is the measurement field. The ratio $B_m/G$ is a curvature radius that describes how straight lines within the object become curved in the MRI image. In our human hand experiment, $L_z \approx L_x \approx$ 14 cm and $L_y \approx$ 4 cm. The parameter $\varepsilon$ for the $G_y$ gradient, $L_y(G_{y,max}/B_m)$=0.04, is thus seven times lower than for the $G_z$ gradient, $L_z(G_{z,max}/B_m)$=0.3. Because the concomitant gradient artifacts associated with $G_z$ are rather small, as shown below, the effect of $G_y$ can be safely neglected. It is sufficient, therefore, to consider a two-dimensional problem in the XZ plane.

Following the analysis of [40], we can write for the gradient echo sequence with $G_x$ and $G_z$ gradients:

$$x_{MRI} \approx x + \frac{G_x}{2B_m} z^2, \qquad (1)$$

$$z_{MRI} \approx \left(1 + \frac{G_x}{2B_m} x\right) \cdot z + \frac{G_z}{8B_m} x^2. \qquad (2)$$

In these formulas, $(x, z)$ are coordinates of a voxel inside an object, $(x_{MRI}, z_{MRI})$ are coordinates of the corresponding pixel in the MRI image, $G_x$ is the readout gradient (after the reversal), and $G_z$ is a given value of the phase encoding gradient. Eq. (1) and the first term in Eq. (2) depend on the constant $G_x$, and describe geometrical deformation of the image. The last term in Eq. (2) depends on the gradient $G_z$ that takes a discrete set of values and causes blurring of the image along the Z direction [40]. In the human hand experiment, the ratio $G_x/2B_m$ is equal to -1.03·10$^{-2}$ cm$^{-1}$. For the maximum value of $|z| \approx$ 7 cm, the difference $x_{MRI}-x$ is about -5 mm. The last term in Eq. (2) is four times smaller even for the largest $G_{z,max}=|G_x|$ and $|x| \approx$ 7 cm, and equals 1.25 mm in that case. Therefore, the magnitude of image blurring in the worst case is less than the pixel size (3 mm). Based on these estimates, we neglect the last term of Eq. (2) in the present work.

The effect of concomitant gradients can then be described as a deformation of an image that makes lines of constant $x$ curved, and lines of constant $z$ – tilted [40]. As a result, the Fourier reconstructed image becomes pointed (i.e. convex and narrowing) in the direction of decreasing readout frequencies. The corrected image can be obtained by transformation of coordinates $(x_{MRI}, z_{MRI})$ to $(x, z)$ according to Eq. (1) and Eq. (2) (without the last term). Results of such correction are presented in Section 3.1.

*2.4 Noise propagation*

Sources of noise and noise propagation in multichannel ULF MRI are very different from those in conventional high-field MRI. Because array coils used for parallel imaging at high fields are inductively decoupled [2,4], noise correlations among the coils arise from electric currents induced inside a sample [2]. Elements of the noise resistance matrix, used to describe this effect, scale with the Larmor frequency as $\omega^2$ [2]. Therefore, noise correlations due to the presence of a sample are negligible at microtesla fields. The main sources of noise in ULF MRI are the SQUIDs themselves and the cryostat. The SQUID noise is Gaussian white noise [31] (the $1/f$ noise in modern low-$T_c$ SQUIDs becomes relevant below 1 Hz). The cryostat noise is Johnson noise produced by conductive components, such as the thermal shield and superinsulation. Noise correlations in multichannel SQUID systems arise from mutual inductive coupling of pick-up coils, and from the fact that the cryostat noise contributions are correlated to some degree for different channels. These sources of noise correlation are studied in Section 3.2.

Noise properties of a multichannel system can be characterized by the noise covariance matrix:

$$\Psi_{ij} = \frac{1}{N} \sum_{m=1}^{N} n_i(m) n_j^*(m) \qquad (3)$$

Here, $n_i(m)$ is a noise signal value of $i$-th channel for measurement $m$, and $N$ is the total number of measurements. The noise correlation matrix is defined as $\Psi_{ij}^{\text{corr}} = \Psi_{ij} / \sqrt{\Psi_{ii} \Psi_{jj}}$.

If the noise covariance matrix and complex coil sensitivities are known, a composite image can be obtained from individual-channel images using the maximum-SNR multicoil reconstruction [2]:

$$M = \frac{S^H \Psi^{-1} I}{S^H \Psi^{-1} S} \qquad (4)$$

In this formula, which is used for each pixel in the image, $I$ is a column vector consisting of pixel values from $n_C$ channels, $S$ is a vector of channel sensitivities at that pixel, $\Psi$ is the $n_C \times n_C$ noise covariance matrix, and $M$ is the resulting pixel value. This reconstruction method performs both sensitivity correction and SNR optimization. If $\Psi$ is replaced with the identity matrix, the method reduces to sensitivity correction.

SNR of a detector array can be quantified as [2]
$$\text{SNR} \sim (S^H \Psi^{-1} S)^{1/2} \qquad (5)$$

This quantity is a measure of local SNR [2], and reflects properties of the array itself. Its inverse is sometimes referred to as the basic array noise [42]. Actual SNR values also depend on other factors such as sequence parameters. Results of applying Eqs. (4)-(5) to our experimental data are presented in Section 3.2.

*2.4 SENSE image reconstruction*

Accelerated image acquisition in our phantom experiment was achieved by undersampling along the phase encoding direction Z. Non-aliased full-FOV images were then reconstructed using SENSE method [6]. In the simplest case of 2D Fourier imaging, this method (1D SENSE) works as follows.

The number of phase encoding steps, and thus the scan time, is reduced by factor $R$, called the reduction (or acceleration) factor. This is accomplished by increasing the increment value for the phase encoding gradient $G_z$ while preserving its limiting values. The spatial resolution remains unchanged, but the imaging FOV along the phase encoding direction is reduced by the factor $R$, which causes aliasing. This means that the image signal at an aliased pixel within the reduced FOV is a superposition of up to $R$ signals corresponding to equidistant pixels in the full FOV. The SENSE image reconstruction from the undersampled data is performed in two steps. First, reduced-FOV aliased images from individual detector coils are obtained via discrete Fourier transform. Second, a non-aliased full-FOV image is reconstructed from the individual images using full-FOV sensitivity maps for all the coils. Because different coils in an array have different local sensitivities, the superposition of $R$ signals due to aliasing occurs with different weights for different coils [6], and can be undone by means of linear algebra if the number of coils is greater than $R$. This allows unfolding of the aliased images.

The unfolding step of the SENSE reconstruction is performed according to the formula [6]:

$$M = [(S^H \Psi^{-1} S)^{-1} S^H \Psi^{-1}] I \qquad (6)$$

As in Eq. (4) above, $\Psi$ is the $n_C \times n_C$ noise covariance matrix of the array, and $I$ is the $n_C$ long column vector, consisting of pixel values provided by the array coils. The matrix $S$ in Eq. (6) is a $n_C \times n_A$ matrix of sensitivity values, where $n_A$ is the number of pixels superposed due to aliasing ($n_A \leq R$). $M$ is a column vector of size $n_A$, that includes values of $n_A$ equidistant pixels after the unfolding. For $R=1$, Eq. (6) reduces to Eq. (4).

The imaging acceleration, however, is achieved at the expense of reduction in signal-to-noise ratio. The SNR at a given pixel after SENSE reconstruction with a certain $R$ is related to the SNR at the same pixel in a fully encoded image as [6]

$$\text{SNR}_R = \frac{\text{SNR}_{\text{Full}}}{g\sqrt{R}} \qquad (7)$$

The factor √R in the denominator results from the R-fold overall reduction in the acquisition time. The factor g (g ≥ 1), called the geometry factor, describes local noise amplification in the unfolded image. It depends on both R and geometry of the sensor array. A simple description of SENSE method can be found, e.g., in [43]. Section 3.3 presents SENSE reconstructed images of the 2D phantom and corresponding g factor maps for different values of R.

## 3. Results

In this Section, we study the benefits of using multiple channels in ULF MRI. The FOV and SNR improvements, as well as effects of concomitant gradients and noise propagation, are studied in the case of the human hand. The imaging acceleration is demonstrated for the 2D water phantom, because SNR in this case is sufficiently high and allows imaging without signal averaging.

### 3.1 FOV and concomitant gradient artifacts

Results of 3D imaging of the human hand at 46 µT measurement field are exhibited in Fig. 3. The experiment was performed as described in Section 2.2. Each image in Fig. 3 represents a 6 mm thick horizontal layer of the hand with the vertical position of that layer's central plane specified by coordinate Y. The imaging resolution in the XZ plane is 3 mm × 3 mm. The coordinates are given with respect to the center of the imaging coil system. The cryostat bottom was located at Y ≈ 25 mm. Each image in Fig. 3 is a composite seven-channel image, computed as a square root of the sum of squares of images from the seven individual channels with the same Y. The composite images were subjected to fine-mesh bicubic interpolation, followed by correction of concomitant gradient artifacts as explained below. The images in Fig. 3 are proton density images with some $T_2$ contrast (because measurement starts about 50 ms after $B_m$ is applied). They show important anatomical features of the hand, including soft tissues, joints, and bones.

Fig. 4 shows seven individual-channel images of the layer with Y = 12 mm. The images are exhibited with different intensity scales to emphasize the extent of each channel's FOV. According to Fig. 4, any single channel in our system can only image a part of the hand. The array of seven channels used simultaneously, however, expands the FOV and allows high quality imaging of the whole hand.

Image distortions caused by concomitant gradients are studied in Fig. 5. This figure exhibits a sensitivity map and a 2D hand image both before (A) and after (B) the concomitant gradient correction. 3D sensitivity maps were obtained for all the channels by imaging a large and thick uniform water phantom. The map in Fig. 5 is a composite seven-channel sensitivity map for Y = 18 mm, corresponding to the top image layer in Fig. 3. The 2D hand image was computed from the 3D data as a square root of the sum of squares, with summation over seven channels and four image layers. The 2D image is considered here for the purpose of comparison with our first 2D hand image [23]. The ratio $G_x/2B_m$ in Eqs. (1) and (2) is equal to $-1.03 \cdot 10^{-2}$ cm$^{-1}$ in this experiment. According to Fig. 5, the sensitivity map before the correction (A) is visibly deformed: it is convex to the right and narrows from the left to the right. This is the direction (from Ch 6 to Ch 3) in which the readout frequency decreases. The hand image (A) exhibits the same distortion, which is similar to the distortion in [23], but with the opposite direction.

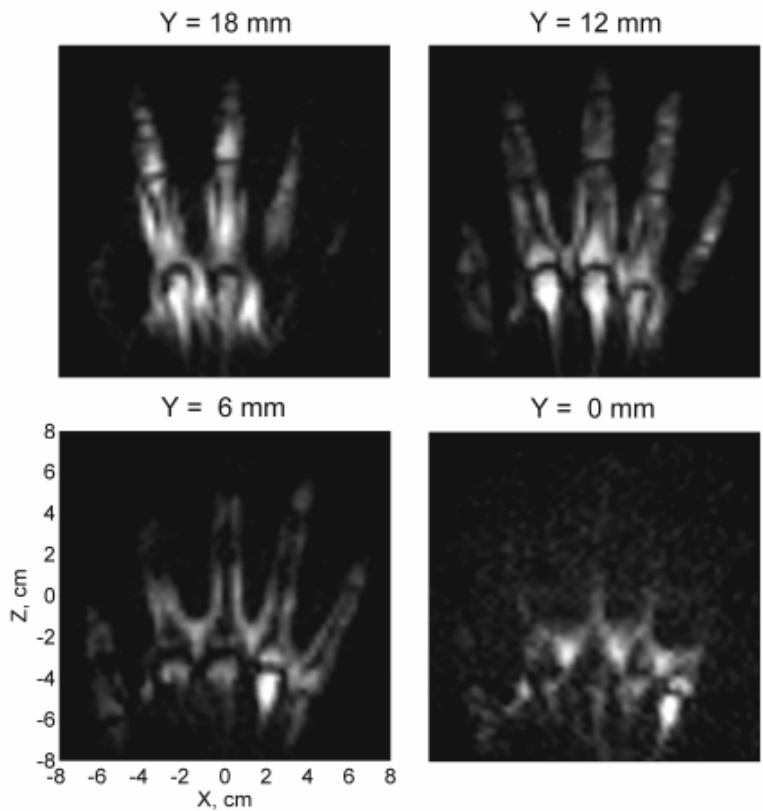

Fig. 3. Composite seven-channel 3D image of the human hand acquired at 46 µT.

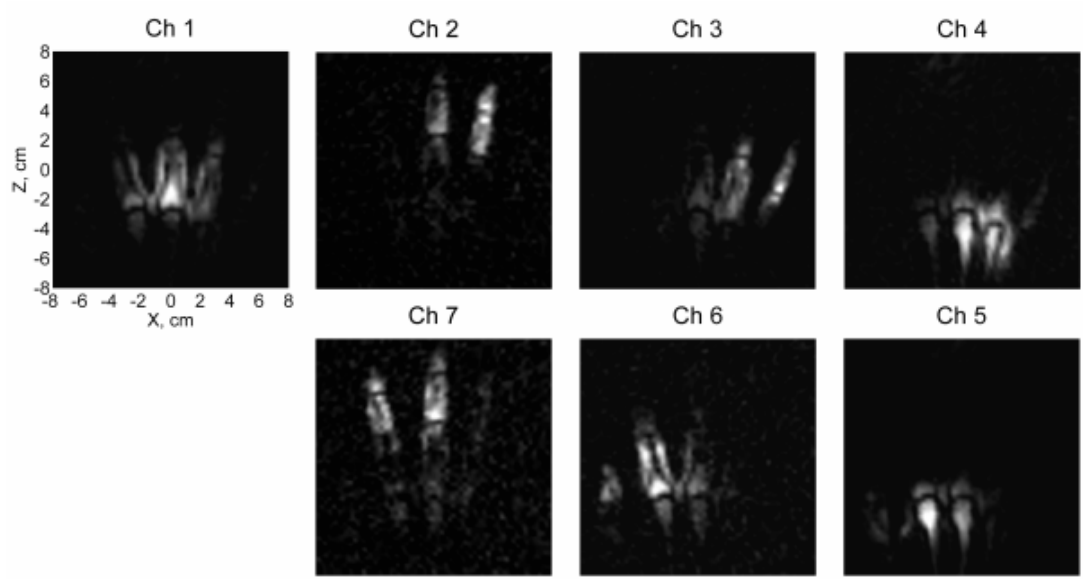

Fig. 4. Images from the seven individual channels for $Y$=12 mm.

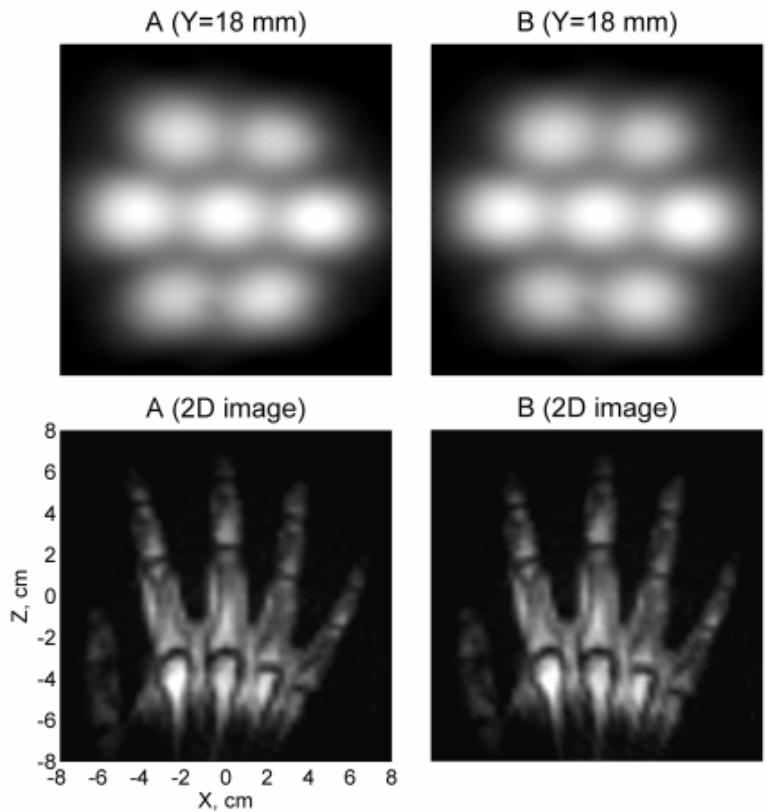

Fig. 5. Images before (A) and after (B) correction of concomitant gradient artifacts.

Correction of concomitant gradient artifacts is performed in two steps. First, the size of the image element is reduced by fine-mesh bicubic interpolation. Then, the image coordinates are transformed from ($x_{MRI}$, $z_{MRI}$) to ($x$, $z$) according to Eq. (1) and Eq. (2) (without the last term). Fig. 5 shows that the sensitivity map becomes more symmetric after such correction (B). It should be noted that the whole sensitivity map is slightly turned clockwise with respect to the X axis because of the imperfect orientation of the cryostat. The corrected image of the hand (B) features straight fingers and does not exhibit any obvious deformations. These results show that geometrical image distortions caused by concomitant gradients are relatively small in our experiments, and can be easily corrected in the final images.

### 3.2 Noise propagation and SNR

The experimentally determined noise covariance matrix for our seven-channel SQUID system is given in Eq. (8). Noise measurements were performed for the total time of 150 s with 16 kHz sampling rate, and the $\Psi$ matrix was computed according to Eq. (3). The matrix elements are given in units of (mV)$^2$. The matrices in Eqs. (8)-(10) are symmetric. The first observation one can make from Eq. (8) is that the noise power for channels 2-7 is higher than for channel 1. This is consistent with the previously reported noise spectra of our system [24]. The higher noise levels for channels 2-7, surrounding channel 1, are due to Johnson noise from the thermal shield (consisting of aluminum rods) between the vertical walls of the cryostat [24]. Each of the seven channels exhibited essentially the same (low) noise level when installed at the center (Ch 1 position).

$$\Psi = \begin{bmatrix} 0.079 & 0.026 & 0.028 & 0.030 & 0.026 & 0.028 & 0.029 \\ & 0.202 & 0.050 & 0.015 & 0.009 & 0.015 & 0.055 \\ & & 0.223 & 0.064 & 0.016 & 0.011 & 0.016 \\ & & & 0.261 & 0.058 & 0.016 & 0.012 \\ & & & & 0.197 & 0.055 & 0.015 \\ & & & & & 0.246 & 0.062 \\ & & & & & & 0.234 \end{bmatrix} \quad (8)$$

The noise correlation matrix, Eq. (9), further explains noise properties of the system. The noise correlation between channel 1 and any of the surrounding channels is about 0.21, while the correlation between any two nearby channels next to the wall of the cryostat is about 0.25. This means that thermal noise propagates from the cryostat walls inwards. The correlations between any two remotely positioned channels are 0.05 – 0.07.

$$\Psi^{corr} = \begin{bmatrix} 1.000 & 0.205 & 0.208 & 0.210 & 0.204 & 0.202 & 0.214 \\ & 1.000 & 0.237 & 0.067 & 0.047 & 0.068 & 0.251 \\ & & 1.000 & 0.265 & 0.074 & 0.045 & 0.071 \\ & & & 1.000 & 0.254 & 0.065 & 0.049 \\ & & & & 1.000 & 0.250 & 0.071 \\ & & & & & 1.000 & 0.257 \\ & & & & & & 1.000 \end{bmatrix} \quad (9)$$

It should be noted that these noise correlation results compare favorably with the noise correlation performance achieved in high-field MRI with phased arrays specially designed to improve SNR. It was shown in the original work on phased arrays [2] that off-diagonal elements of the electric coupling matrix, which determine noise correlation, can be as large as 0.41. It was demonstrated in the same work that, if noise correlations are completely ignored, the maximum penalty in terms of SNR is about 10% [2]. For this reason, noise correlations were neglected in many applications of phased arrays [4].

It might be interesting to consider how the noise correlation levels will change if the cryostat noise is reduced or eliminated. In this case, propagation of SQUID noise due to inductive coupling of pick-up coils will, presumably, be the main source of correlations. In the first approximation, the fraction of magnetic flux in one SQUID channel that penetrates into another is equal to $M_p/(L_p+L_i)$, where $M_p$ is the mutual inductance of two pick-up coils, $L_p$ is inductance of one pick-up coil, and $L_i$ is inductance of the SQUID input coil (in series with the pick-up coil). We calculated the magnetic coupling matrix $k_m$ for our system of seven SQUID gradiometers. The matrix, Eq. (10), shows that magnetic flux coupling, and, therefore, noise correlation is around 0.016 for nearby channels, and of the order of 0.002 for remote channels. To check these estimates experimentally, we measured small-signal crosstalk of the channels. A small 1940 Hz sinewave test signal corresponding to 1 $\Phi_0$ peak-to-peak was applied to each SQUID, one channel at a time.

$$k_m = \begin{bmatrix} 1.000 & 0.014 & 0.015 & 0.017 & 0.015 & 0.017 & 0.018 \\ & 1.000 & 0.013 & 0.002 & 0.001 & 0.002 & 0.015 \\ & & 1.000 & 0.016 & 0.002 & 0.001 & 0.002 \\ & & & 1.000 & 0.016 & 0.002 & 0.001 \\ & & & & 1.000 & 0.016 & 0.002 \\ & & & & & 1.000 & 0.018 \\ & & & & & & 1.000 \end{bmatrix} \quad (10)$$

Amplitudes of signals at the same frequency, measured by the seven channels in the flux-locked mode, were then compared. The channels' crosstalk was always below 1%, in a general agreement with the computational estimates. These results suggest that noise correlation in multichannel ULF MRI can be much lower than in high-field MRI, provided that special low-noise cryostats are used.

Application of the maximum-SNR multicoil image reconstruction method is illustrated in Fig. 6A. The hand image in this figure was computed according to Eq. (4), with the noise covariance matrix, Eq. (8), taken into account. However, when Eq. (4) was used with $\Psi$ set to the identity matrix, a very similar image was obtained (not shown). This fact suggests that the effect of noise correlations in our system on the quality of multichannel image reconstruction is indeed small. Fig. 6B exhibits a map of the relative SNR, Eq. (5), with the maximum SNR value set to 10. Unlike the composite sensitivity map in Fig. 5, the map in Fig. 6B clearly shows that the SNR for channel 1 is about twice as high as that for the surrounding channels.

### 3.3 Demonstration of imaging acceleration

Results of the 2D phantom imaging experiment are presented in Fig. 7. The imaging was performed with 2 mm × 2 mm resolution according to the procedure described in Section 2.2. Because the imaging gradients were relatively weak in this experiment, no correction of concomitant gradient artifacts was employed. The top row of images in Fig. 7 contains composite (square root of the sum of squares) seven-channel images acquired with different degrees of undersampling along the phase encoding direction Z characterized by the reduction factor $R$. Single-average acquisition of the full-FOV image with $R=1$ included 73 phase encoding steps and required 6 min. For $R>1$, these values were reduced $R$ times, as explained in Section 2.5. The number of phase encoding steps, taken symmetrically with respect to $G_z=0$, was $2[36/R]+1$, and the total imaging time was approximately $6/R$ min. The accelerated image acquisition based on such undersampling led to reduced FOV and aliasing, as seen in the composite images with $R=2,3,4$. The middle row in Fig. 7 exhibits corresponding full-FOV images reconstructed from the undersampled data using 1D SENSE method with experimentally determined sensitivity maps of the sensors. The identity matrix was used instead of $\Psi$ in Eq. (6), but this did not significantly affect the results. The composite seven-channel 2D sensitivity map is shown as the first image in the bottom row of Fig. 7. The other images in that row are maps of the geometry factor for $R=2,3$, and 4.

According to Fig. 7, the SENSE reconstruction provides nearly perfect phantom images for $R=1$, 2, and 3. In the case of $R=1$, SENSE method reduces to sensitivity correction, Eq. (4), without unfolding. The corrected image shows more uniform intensity distribution than the composite image with $R=1$. The SENSE algorithm allows successful unfolding of the aliased images with $R=2$ and $R=3$, enabling accelerated image acquisition. The SENSE image with $R=4$, however, is marked by excessive noise. This result is consistent with behavior of the geometry factor, that quantitatively describes local noise amplification in the unfolded images. While the maximum $g$ factor values for $R=2$ and $R=3$ are 1.20 and 2.06, respectively, the maximum value for $R=4$ is as high as 47. This means that the SENSE reconstruction becomes increasingly unstable for $R\geq4$, suggesting that the spatial information provided by the sensor array is no longer sufficient for unfolding the aliased images.

This limitation can be easily understood if one considers the sensitivity map in Fig. 7. Because the pick-up coils in our system form a two-dimensional array and have substantially localized sensitivities, any straight line parallel to the Z axis intersects essential sensitivity regions of at most three coils. Thus, no more than three coils can efficiently participate in 1D SENSE reconstruction for any given set of pixels related by aliasing. As a result, the maximum practical acceleration of *R*=3 is lower than the number of channels in our system. Higher accelerations can be achieved in 3D Fourier imaging by undersampling along each of the two phase encoding directions and reconstructing images with 2D SENSE [44]. This approach, however, is not very practical with our present system, because it would require phase encoding to be performed along X, and frequency encoding – along Y direction, which is inefficient due to the limited sensitivity depth of our sensors.

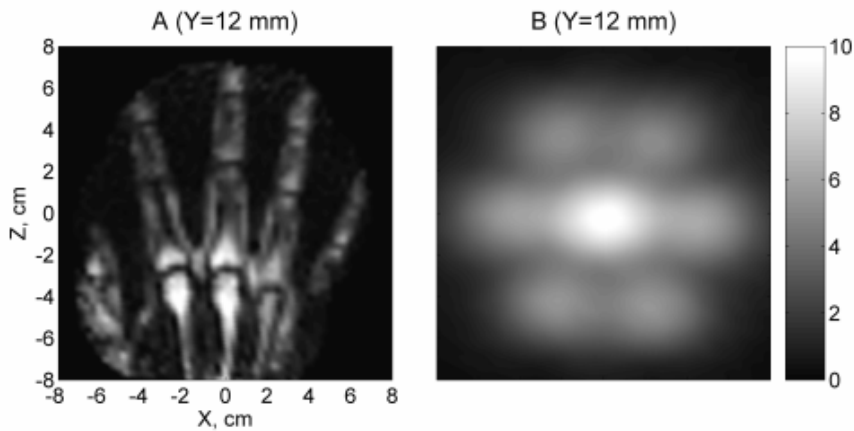

Fig. 6. A) Image obtained using the maximum-SNR multichannel reconstruction method, taking into account noise propagation. B) Relative SNR of the seven-channel system, determined from sensitivity maps and noise matrix.

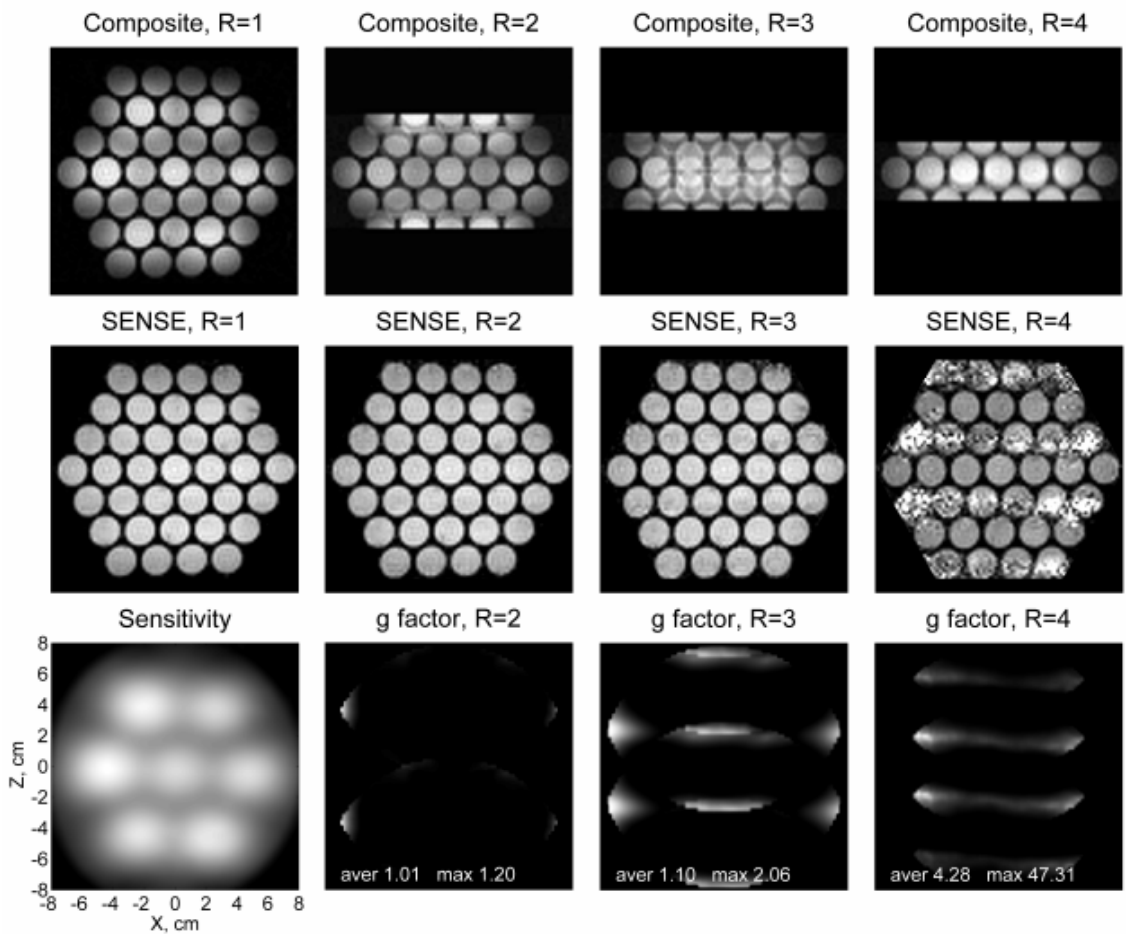

Fig. 7. SENSE reconstruction of 2D phantom images for different values of the acceleration factor *R*.

It should be emphasized that the results in Fig. 7 demonstrate real imaging acceleration for the 2D water phantom. The SENSE images with *R*=2 and *R*=3 were acquired in 3 min and 2 min, respectively, while the single-average acquisition of the fully encoded image (*R*=1) took 6 min. The SENSE reconstruction, illustrated in Fig. 7, can also be applied to the human hand images, reported in this paper. However, anatomical imaging with our present system requires signal averaging, so no actual imaging acceleration can be achieved in vivo at this time.

## 4. Conclusion

In this work, we used our seven-channel SQUID system for 3D ULF MRI and MEG to experimentally study parallel imaging at microtesla fields. We showed that image acquisition with the sensor array instead of a single sensor substantially improves FOV and SNR in 3D anatomical imaging. Moreover, 3-fold imaging acceleration based on SENSE method was demonstrated at ULF for the first time. These results indicate that parallel imaging techniques are efficient in ULF MRI, and can significantly enhance performance of multichannel SQUID instruments. As in high-field MRI, accelerated image acquisition is the most promising application of sensor arrays at ULF. The intrinsic SNR of our present system, however, is not high enough to allow accelerated imaging in vivo. Thus, the FOV and SNR improvements by the sensor array remain the main advantages of multichannel ULF MRI. By using stronger pre-polarizing fields and reducing the cryostat noise, we should be able to increase the system SNR sufficiently to achieve imaging acceleration in human subject experiments.

**Acknowledgments**

We gratefully acknowledge the support of the U.S. National Institutes of Health Grant R01-EB006456 and of the Los Alamos National Security, LLC, for the National Nuclear Security Administration of the U.S. Department of Energy Grant LDRD-20060312ER.

**References**

[1] J.S. Hyde, A. Jesmanowicz, W. Froncisz, J.B. Kneeland, T.M. Grist, N.F. Campagna, Parallel image acquisition from noninteracting local coils, J. Magn. Reson. 70 (1986) 512-517.
[2] P.B. Roemer, W.A. Edelstein, C.E. Hayes, S.P. Souza, O.M. Mueller, The NMR phased array, Magn. Reson. Med. 16 (1990), 192-225.
[3] L.L. Wald, L. Carvajal, S.E. Moyher, S.J. Nelson, P.E. Grant, A.J. Barkovich, D.B. Vigneron, Phased array detectors and an automated intensity-correction algorithm for high-resolution MR imaging of the human brain, Magn. Reson. Med. 34 (1995) 433-439.
[4] S.M. Wright, L.L. Wald, Theory and application of array coils in MR spectroscopy, NMR Biomed. 10 (1997) 394-410.
[5] D.K. Sodickson, W.J. Manning, Simultaneous acquisition of spatial harmonics (SMASH): fast imaging with radiofrequency coil arrays, Magn. Reson. Med. 38 (1997) 591-603.
[6] K.P. Pruessmann, M. Weiger, M.B. Scheidegger, P. Boesiger, SENSE: sensitivity encoding for fast MRI, Magn. Reson. Med. 42 (1999) 952-962.
[7] K.P. Pruessmann, M. Weiger, P. Bornert, P. Boesiger, Advances in sensitivity encoding with arbitrary $k$-space trajectories, Magn. Reson. Med. 46 (2001) 638-651.
[8] D.K. Sodickson, C.A. McKenzie, A generalized approach to parallel magnetic resonance imaging, Med. Phys. 28 (2001) 1629-1643.
[9] K.P. Pruessmann, Encoding and reconstruction in parallel MRI, NMR Biomed. 19 (2006) 288-299.
[10] T. Niendorf, D.K. Sodickson, Parallel imaging in cardiovascular MRI: methods and applications, NMR Biomed. 19 (2006) 325-341.
[11] Y. Zhu, C.J. Hardy, D.K. Sodickson, R.O. Giaquinto, C.L. Dumoulin, G. Kenwood, T. Niendorf, H. Lejay, C.A. McKenzie, M.A. Ohliger, N.M. Rofsky, Highly parallel volumetric imaging with a 32-element RF coil array, Magn. Reson. Med. 52 (2004) 869-877.
[12] J.A. de Zwart, P. van Gelderen, X. Golay, V.N. Ikonomidou, J.H. Duyn, Accelerated parallel imaging for functional imaging of the human brain, NMR Biomed. 19 (2006) 342-351.
[13] M.P. McDougall, S.M. Wright, 64-channel array coil for single echo acquisition magnetic resonance imaging, Magn. Reson. Med. 54 (2005) 386-392.
[14] F.H. Lin, L.L. Wald, S.P. Ahlfors, M.S. Hämäläinen, K.K. Kwong, J.W. Belliveau, Dynamic magnetic resonance inverse imaging of human brain function, Magn. Reson. Med. 56 (2006) 787-802.
[15] M. Hämäläinen, R. Hari, R. Ilmoniemi, J. Knuutila, O. Lounasmaa, Magnetoencephalography – theory, instrumentation, and application to non-invasive studies of the working human brain, Rev. Mod. Phys. 65 (1993) 413-497.
[16] R. McDermott, A.H. Trabesinger, M. Mück, E.L. Hahn, A. Pines, J. Clarke, Liquid-state NMR and scalar couplings in microtesla magnetic fields, Science 295 (2002) 2247-2249.
[17] R. McDermott, S.K. Lee, B. ten Haken, A.H. Trabesinger, A. Pines, J. Clarke, Microtesla MRI with a superconducting quantum interference device, Proc. Nat. Acad. Sci. 101 (2004) 7857-7861.
[18] R. McDermott, N. Kelso, S.K. Lee, M. Mößle, M. Mück, W. Myers, B. ten Haken, H.C. Seton, A.H. Trabesinger, A. Pines, J. Clarke, SQUID-detected magnetic resonance imaging in microtesla magnetic fields, J. Low Temp. Phys. 135 (2004) 793-821.
[19] M. Mößle, W.R. Myers, S.K. Lee, N. Kelso, M. Hatridge, A. Pines, J. Clarke, SQUID-detected in vivo MRI at microtesla magnetic fields, IEEE Trans. Appl. Supercond. 15 (2005) 757-760.
[20] A.N. Matlachov, P.L. Volegov, M.A. Espy, J.S. George, R.H. Kraus, Jr., SQUID detected NMR in microtesla magnetic fields, J. Magn. Reson. 170 (2004) 1-7.
[21] P. Volegov, A.N. Matlachov, M.A. Espy, J.S. George, R.H. Kraus, Jr., Simultaneous magnetoencephalography and SQUID detected nuclear MR in microtesla magnetic fields, Magn. Reson. Med. 52 (2004) 467-470.
[22] M.A. Espy, A.N. Matlachov, P.L. Volegov, J.C. Mosher, R.H. Kraus, Jr., SQUID-based simultaneous detection of NMR and biomagnetic signals at ultra-low magnetic fields, IEEE Trans. Appl. Supercond. 15 (2005) 635-639.
[23] V.S. Zotev, A.N. Matlachov, P.L. Volegov, H.J. Sandin, M.A. Espy, J.C. Mosher, A.V. Urbaitis, S.G. Newman, R.H. Kraus, Jr., Multi-channel SQUID system for MEG and ultra-low-field MRI, IEEE Trans. Appl. Supercond. 17 (2007) 839-842.
[24] V.S. Zotev, A.N. Matlashov, P.L. Volegov, A.V. Urbaitis, M.A. Espy, R.H. Kraus, Jr., SQUID-based instrumentation for ultralow-field MRI, Supercond. Sci. Technol. 20 (2007) S367-S373.
[25] V.S. Zotev, A.N. Matlashov, P.L. Volegov, I.M. Savukov, M.A. Espy, J.C. Mosher, J.J. Gomez, R.H.Kraus, Jr., Microtesla MRI of the human brain combined with MEG, submitted, preprint available at http://arxiv.org/abs/0711.0222
[26] M. Burghoff, S. Hartwig, L. Trahms, Nuclear magnetic resonance in the nanotesla range, Appl. Phys. Lett. 87 (2005) 054103.
[27] M. Burghoff, S. Hartwig, W. Kilian, A. Vorwerk, L. Trahms, Thiel F, Hartwig S, Trahms L, SQUID systems adapted to record nuclear magnetism in low magnetic fields, IEEE Trans. Appl. Supercond. 17 (2007) 846-849.
[28] S.K. Lee, M. Mößle, W. Myers, N. Kelso, A.H. Trabesinger, A. Pines, J. Clarke, SQUID-detected MRI at 132 µT with $T_1$-weighted contrast established at 10 µT-300mT, Magn. Reson. Med. 53 (2005) 9-14.
[29] M. Mößle, S.I. Han, W.R. Myers, S.K. Lee, N. Kelso, M. Hatridge, A. Pines, J. Clarke, SQUID-detected microtesla MRI in the presence of metal, J. Magn. Reson. 179 (2006) 146-151.
[30] A. Macovski, S. Conolly, Novel approaches to low cost MRI, Magn. Reson. Med. 30 (1993) 221-230.
[31] J. Clarke, A.I. Braginski (Eds.), The SQUID Handbook, Wiley-VCH, Weinheim, 2004.
[32] N.Q. Fan, M.B. Heaney, J. Clarke, D. Newitt, L.L. Wald, E.L. Hahn, A. Bielecki, A. Pines, Nuclear magnetic resonance with dc SQUID preamplifiers, IEEE Trans. Magn. 25 (1989) 1193-1199.
[33] H.C. Seton, J.M.S. Hutchison, D.M. Bussel, A 4.2K receiver coil and SQUID amplifier used to improve the SNR of low-field magnetic resonance images of the human arm, Meas. Sci. Technol. 8 (1997) 198-207.
[34] K. Schlenga, R. McDermott, J. Clarke, R.E. de Souza, A. Wong-Foy, A. Pines, Low-field magnetic resonance imaging with a high-$T_c$ dc superconducting quantum interference device, Appl. Phys. Lett. 75 (1999) 3695-3697.
[35] Ya.S. Greenberg, Application of superconducting quantum interference devices to nuclear magnetic resonance, Rev. Mod. Phys. 70 (1998) 175-222.
[36] J. Stepišnik, V. Eržen, M. Kos, NMR imaging in the Earth's

magnetic field, Magn. Reson. Med. 15 (1990) 386-391.

[37] A. Mohorič, G. Planinšič, M. Kos, A. Duh, J. Stepišnik, Magnetic resonance imaging system based on Earth's magnetic field, Instrum. Sci. Technol. 32 (2004) 655-667.

[38] G. Stroink, W. Moshage, S. Achenbach, Cardiomagnetism, in: W. Andrä, H. Nowak (Eds.), Magnetism in Medicine, Wiley-VCH, Berlin, 1998, pp. 136-189.

[39] P.L. Volegov, J.C. Mosher, M.A. Espy, R.H. Kraus, Jr., On concomitant gradients in low-field MRI. J. Magn. Reson. 175 (2005) 103-113.

[40] W.R. Myers, M. Mößle, J. Clarke, Correction of concomitant gradient artifacts in experimental microtesla MRI, J. Magn. Reson. 177 (2005) 274-284.

[41] D.A. Yablonskiy, A.L. Sukstanskii, J.J.H. Ackerman, Image artifacts in very low magnetic field MRI: the role of concomitant gradients, J. Magn. Reson. 174 (2005) 279-286.

[42] M. Weiger, K.P. Pruessmann, C. Leussler, P. Röschmann, P. Boesiger, Specific coil design for SENSE: a six-element cardiac array, Magn. Reson. Med. 45 (2001) 495-504.

[43] M.A. Bernstein, K.F. King, X.J. Zhou, Handbook of MRI Pulse Sequences, Elsevier Academic Press, 2004, pp. 527-531.

[44] M. Weiger, K.P. Pruessmann, P. Boesiger, 2D SENSE for faster 3D MRI, MAGMA 14 (2001)10-19.